\documentclass[prl,aps,amssymb,twocolumn,reprint,superscriptaddress]{revtex4-1}
\usepackage{xeCJK}
\usepackage{amsmath}
\usepackage{amssymb}
\usepackage{amsthm}
\usepackage{amsfonts}
\usepackage{listings}
\usepackage{enumerate}
\usepackage{latexsym}
\usepackage{color}
\usepackage{xcolor}
\usepackage{bm}
\usepackage{hyperref}
\hypersetup{
 pdfnewwindow=true, colorlinks=true,
 linkcolor=blue, anchorcolor=blue,
 citecolor=blue, filecolor=blue,
 menucolor=blue, urlcolor=blue}

\usepackage{psfrag}

\usepackage{graphicx}
\usepackage{subfigure}
\usepackage{lipsum}
\usepackage{float}
\usepackage{braket}
\usepackage{array}
\usepackage{soul} 
\usepackage{titlesec}

\newcommand{\beginsupplement}{%
        \setcounter{table}{0}
        \renewcommand{\thetable}{S\arabic{table}}%
        \setcounter{figure}{0}
        \renewcommand{\thefigure}{S\arabic{figure}}%
        \setcounter{equation}{0}
        \renewcommand{\theequation}{S\arabic{equation}}
     }
     
\begin{document}
\soulregister{\cite}7 
\soulregister{\ref}7
\tolerance 10000

\draft

\title{Percolation from Quantum Metric in Flat-Band Delocalization}


\author{Bo Yin}
\affiliation{Beijing National Laboratory for Condensed Matter Physics,
and Institute of Physics, Chinese Academy of Sciences, Beijing 100190, China}
\affiliation{University of Chinese Academy of Sciences, Beijing 100049, China}

\author{Zhijun Wang}
\email{wzj@iphy.ac.cn}
\affiliation{Beijing National Laboratory for Condensed Matter Physics,
and Institute of Physics, Chinese Academy of Sciences, Beijing 100190, China}
\affiliation{University of Chinese Academy of Sciences, Beijing 100049, China}

\author{Quansheng Wu}
\email{quansheng.wu@iphy.ac.cn}
\affiliation{Beijing National Laboratory for Condensed Matter Physics,
and Institute of Physics, Chinese Academy of Sciences, Beijing 100190, China}
\affiliation{University of Chinese Academy of Sciences, Beijing 100049, China}

\begin{abstract}
 The quantum metric is a fundamental ingredient of band quantum geometry and has recently attracted intense interest, with most of its transport signatures appearing in the intrinsic second order nonlinear conductivity. In the clean limit, previous works argued that linear response conductivity is insensitive to the quantum metric, while the Berry curvature yields an intrinsic anomalous Hall contribution. Here we combine analytic derivations with new numerics to show that disorder modifies the linear response conductivity dominated by geometric conductivity which is determined by the real space quantum metric. Focusing on a two dimensional multi-flatband stub-pyrochlore lattice, we identify a critical delocalized regime sandwiched between flat band localization and Anderson localization, characterized by finite geometric conductivity. Upon including spin orbit coupling, this regime evolves into a diffusive metallic phase, constituting a two dimensional inverse Anderson transition. Moreover, exploiting the connection between the real space quantum metric marker and the Wannier function spread, we construct a bond-percolation model on a square lattice. The resulting percolation region quantitatively coincides the critical delocalized regime, the exponent of which supports a classical percolation universality class. These findings suggest that flat band delocalization can be understood as a classical percolation of quantum metric puddles. This advances our understanding of quantum geometric contributions to transport and establishes linear response measurements as a new avenue for accessing the quantum metric.

\end{abstract}

\maketitle
The quantum metric provides a gauge invariant measure of the “distance” between neighboring Bloch states in Hilbert space. It constitutes the real part of the quantum geometric tensor\cite{Provost1980Riemannian,Resta2011EurPhysJ} of which the imaginary part is the Berry curvature\cite{Berry1984Quantal,Simon1983Holonomy,Aharonov1987Phase}. Beyond its formal definition, the quantum metric has become a unifying ingredient across  disparate phenomena\cite{Torma2022NatRevPhysQuantumGeometry,Yu2025QuantumGeometryNPJQM,10.1093/nsr/nwae334}: it contributions to the superfluid weight in flat band superconductors\cite{Peotta2015NatCommun,PhysRevLett.117.045303,PhysRevB.95.024515,PhysRevB.102.201112,PhysRevLett.128.087002}, in strongly correlated flat band settings it supplies geometric structure that constrains and stabilizes fractional Chern insulator\cite{PhysRevB.90.165139,PhysRevB.104.045104,PhysRevB.108.205144}. 

In transport, a key advance is that the quantum metric can drive intrinsic nonlinear conductivities\cite{PhysRevLett.132.026301,PhysRevLett.127.277201,PhysRevLett.127.277202} that are not controlled by a relaxation time, making it an experimentally attractive quantum geometric channel complementary to Berry curvature mechanisms. Such intrinsic quantum metric signatures are particularly transparent in antiferromagnets\cite{Gao2023QuantumMetricScience,Wang2023QuantumMetricNature} that break inversion P and time-reversal T individually while preserving the combined PT symmetry which enforces vanishing Berry curvature\cite{PhysRevLett.115.216806}. However, previous theoretical studies\cite{PhysRevB.108.155108,Verma2025PNAS} have shown that in the clean limit, the linear conductivity receives no contribution from the quantum metric, with only the Berry curvature contributing to the anomalous Hall conductance\cite{RevModPhys.82.1959}.

The quantum metric can become the leading geometric scale controlling measurable response coefficients\cite{PhysRevLett.132.026002} in flat band systems, where conventional dispersion driven channels are strongly suppressed. There are two types of flat bands generated by destructive interference\cite{PhysRevB.99.045107}. The first type, observed in systems such as 2D Lieb, 2D pyrochlore, and Kagome lattices, forces the flat band to touch the edge of a dispersive band, resulting in “protected” crossings\cite{PhysRevB.78.125104} due to the real-space topology of the Compact Localized States (CLS)\cite{PhysRevB.34.5208,PhysRevB.95.115135,PhysRevB.97.035161,PhysRevB.54.R17296}. The second type features flat bands whose position relative to dispersive bands is tunable, as in stub and cross-stitch lattices. The flat band in the first case shares similar localization properties with the dispersive bands, whereas in the second case, it differs, indicating that novel transport in flat bands can only be realized in such systems\cite{Leykam2017EPJB}.

Many works have investigated the localization properties in flat band systems induced by destructive interference\cite{PhysRevB.88.224203,PhysRevLett.113.236403,PhysRevA.87.061803,PhysRevB.106.205119,Hou2025FateDisorderTBG}, especially the theoretical and experimental discovery of inverse Anderson transitions\cite{PhysRevLett.96.126401,PhysRevLett.129.220403,PhysRevLett.130.206401}. Recent theoretical studies have linked diffusion transport in one dimensional flat band and twsit bilayer graphene to quantum metric\cite{PhysRevLett.134.126301,Chau2024DisorderInducedDiffusion}. Notably, both theoretical predictions and experiments on superconducting qubit arrays reveal disorder induced delocalization in multi-flat band systems, with a transition from flat band localization to Anderson localization\cite{PhysRev.109.1492,RevModPhys.80.1355,y5kd-7prs}, whereas dispersive bands do not show this phenomenon\cite{PhysRevX.15.021091}. This observation highlights the necessity of studying novel transport properties in multi-flat band systems affected by destructive interference, in order to understand the influence of quantum metric.

In this Letter, we demonstrate that flat band delocalization with linear conductivity is attributed to a classical percolation process governed by the quantum metric. We present a detailed analytic derivation of a linear geometric conductance formula that captures the quantum metric contribution in disordered flat bands. It establishes a direct connection between transport and real space quantum metric\cite{PhysRevB.84.241106}. We then adopt a two dimensional multi-flatband stub-pyrochlore model designed to exclude mechanisms other than the quantum metric, and introduce disorder in the energy window between two flat bands. With spin orbit coupling (SOC) and without SOC, both the purely geometric conductance and the conductance computed from transport simulations are obtained. Finally, by using real space quantum metric marker as a component, we construct a classical geometric percolation model that accounts for the emergence of the geometric conductance.

The real space quantum metric \cite{PhysRevB.107.205133,PhysRevLett.122.166602,PhysRevLett.133.026002,dzqm-9kwj,w761-8nf7,qscv-qxqt,PhysRevB.111.134201} restricted to the occupied states is:
\begin{equation}
\hat g_{\mu\nu}
\;\equiv\;
\pi Re\Big(\hat P\,\hat r_\mu\,\hat Q\,\hat r_\nu\,\hat P
+\hat P\,\hat r_\nu\,\hat Q\,\hat r_\mu\,\hat P\Big),
\label{eq:g_operator_projector}
\end{equation}
where $\hat P$ and $\hat Q$ are occupied and unoccupied projectors, $\hat r_\mu$ is the position operator, all in the band basis.

We define the energy difference Lorentzian distribution in the band basis as
\begin{equation}
L_{mn}(\omega)
\;\equiv\;
\frac{2\eta \omega_{mn}}{\omega_{mn}^2+\eta^2},
\qquad (m\neq n),
\label{eq:Lmn_def}
\end{equation}
with $L_{nn}(\omega)=0$, where $\eta$ is the band broadening, vanishes in the clean limit, $\omega_{mn}$ is the energy difference between band $m$ and $n$.

The geometric conductance is,
\begin{align}
\sigma_{\mu\nu}^{\rm geo} &=\frac{e^2}{Vh}
\mathrm{Tr}\!\left(g_{\mu\nu}\,L\right),
\label{eq:trace_gL}
\end{align}
detailed derivations are provided in the supplementary material. To exclude contribution from the Drude weight\cite{PhysRev.133.A171,PhysRevB.47.7995,PhysRevB.110.L241111}, we computed, in real space under open boundary conditions\cite{PhysRevB.102.205123}, the low frequency Drude weight contribution to the conductivity for different disorder strengths, and found that all of them vanish in the dc limit. The detailed results are presented in the Supplementary Material.
\begin{figure}[htb]
  \centering
  \includegraphics[width=\columnwidth]{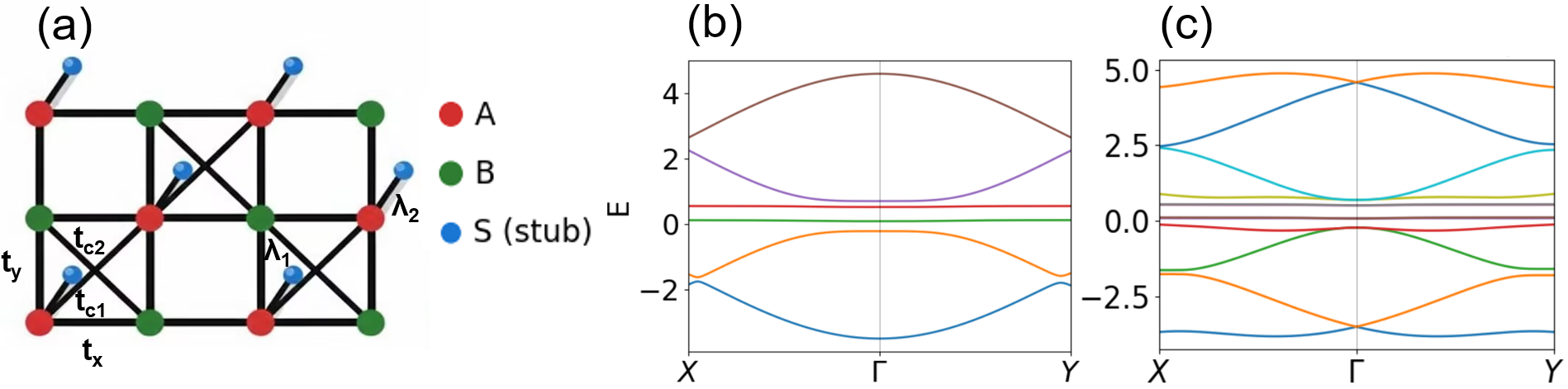}
\caption{2D stub-pyrochlore lattice. (a) Schematic of the two dimensional multi-flatband stub--pyrochlore model on a square lattice, illustrating the hopping processes \(t_x,t_y\), the crossing couplings \(t_{c1},t_{c2}\), and the stub hybridizations \(\lambda_x,\lambda_y\). (b) Band structure of the SOC-free Hamiltonian along the high symmetry path \(X\!-\!\Gamma\!-\!Y\). (c) Band structure of the corresponding spinful Hamiltonian with Rashba SOC \(H_R\) along the same path.}\label{fig:model}
\end{figure}

Previous studies\cite{PhysRevB.103.075415,PhysRevB.105.085154,PhysRevB.106.165133} have suggested that a finite $\eta$, potentially arising from disorder or finite temperature effects, can induce nontrivial transport features in flat bands. However, these works remain largely qualitative and lack quantitative calculations of $\eta$ with concrete numerical values. Here, using an effective medium formulation within the self consistent Born approximation\cite{PhysRevLett.103.196805}, we attribute the disorder contribution to a self consistently determined self energy,
\begin{gather}
\Sigma=\frac{W^2}{12} \frac{a^2}{4 \pi^2} \int_{B Z} d^2 \mathbf{k}  \frac{1}{E_f+i \eta-H(\mathbf{k})-\Sigma} 
\end{gather}
where $W$ is the disorder strength,  a randomon-site disorder potential is uniformly distributed in the range $[-\frac{W}{2},\frac{W}{2}]$. The imaginary part of the self energy components directly yields the disorder induced spectral broadening\cite{PhysRevB.106.134201,PhysRevB.109.035303}, $\eta \sim -\frac{1} {N_b}\,\operatorname{Im}tr\Sigma^{R}$, $N_b$ is the number of bands.

To obtain flat band transport behavior distinct from that of dispersive bands and to extract an unambiguous conductance contribution driven purely by the quantum metric, we employ a model with two CLSs in a 2D stub-pyrochlore lattice\cite{Neves2024CrystalNetCatalog} in Fig.1(a), where the square has side length 0.5. The Hamiltonian is defined by:
\begin{align}
H(\mathbf{k}) &=
\begin{pmatrix}
\epsilon_{A_1} & -t_x f_x & -\lambda_1 & t_{c1} & -t_y f_y & 0\\
-t_x f_x & \epsilon_{B_1} & 0 & -t_y f_y & t_{c2} & 0\\
-\lambda_1 & 0 & \epsilon_{S_1} & 0 & 0 & 0\\
t_{c1} & -t_y f_y & 0 & \epsilon_{A_2} & -t_x f_x & -\lambda_2\\
-t_y f_y & t_{c2} & 0 & -t_x f_x & \epsilon_{B_2} & 0\\
0 & 0 & 0 & -\lambda_2 & 0 & \epsilon_{S_2}
\end{pmatrix}.
\end{align}
where $f_x = 2\cos\!\left(\frac{k_x}{2}\right),f_y = 2\cos\!\left(\frac{k_y}{2}\right),t_x = 1.0, t_y = 1.0,$ $\epsilon_{S_1}$ = 0.546, $\epsilon_{S_2}$ = 0.115, $\epsilon_{A_1}$ = 0.808, $\epsilon_{A_2}$ = 0.927, $\lambda_1$ = 0.102, $\lambda_2$ = 0.191, $\epsilon_{B_1}$ = -0.262, $\epsilon_{B_2}$ = 0.014, $t_{c1}$ = 0.239, $t_{c2}$ = 0.100. The CLSs hybridize and open a trivial gap. Further details on the Hamiltonian construction and parameter generation are provided in the supplementary material. In Fig. 1(b), we present the high symmetry line band structure of the model, where two flat bands, arising from destructive interference, are observed at \(E_f = 0.1\) and \(E_f = 0.54\), well separated from the dispersive bands among the six energy bands. In Fig. 1(c), we incorporate the spin component and adding a Rashba SOC\cite{Rashba1960ExtremumLoop,BychkovRashba1984} term, 
\begin{align}
H_{\mathrm R}(k)
= -2\alpha_{\mathrm R}\big(\sin k_y\,\sigma_x - \sin k_x\,\sigma_y\big)
\end{align}
We choose $\alpha_{\mathrm R}=0.5$. The flat bands at \(E_f = 0.1\) and \(E_f = 0.54\) remain unaffected.

Then we investigate the transport between the two flat bands at \(E_f = 0.3\).
The system consists of an \(L \times L\) sample connected to two conducting leads. To reduce contact resistance, the leads are chosen to have the same Hamiltonian as the sample, but with the Fermi energy \(E_f = -0.5\), placing the leads in the conducting regime. We examine the effects of onsite disorder on the square lattice, where no disorder on the stub. The disorder is modeled by randomly choosing values from a uniform distribution over the range \([-W/2, W/2]\), with \(W\) being the disorder strength and is averaged over 2000 samples. The two-terminal conductance \(G\) is computed using the Landauer–Buttiker formula\cite{Landauer1970,Buttiker1988,FisherLee1981,Datta1995ElectronicTransport} and the recursive Green's function method\cite{MacKinnon1985ZPB,Sancho1985JPhysF}.  
\begin{gather}
G=\frac{e^2}{h}\operatorname{Tr}\left[\Gamma_{\mathrm{l}} \mathrm{G}^{\mathrm{r}} \Gamma_{\mathrm{r}} \mathrm{G}^{\mathrm{a}}\right],
\end{gather}
where $\Gamma_{\mathrm{l}},\Gamma_{\mathrm{r}}$ are the line width function describing the coupling to the left (l) or right (r) lead.
$\mathrm{G}^{\mathrm{r}},\mathrm{G}^{\mathrm{a}}$ are the retarded and advanced Green's function of the scattering region. The trace accounts
for contributions from all transport channels.
\begin{figure}[htb]
  \centering
  \includegraphics[width=\columnwidth]{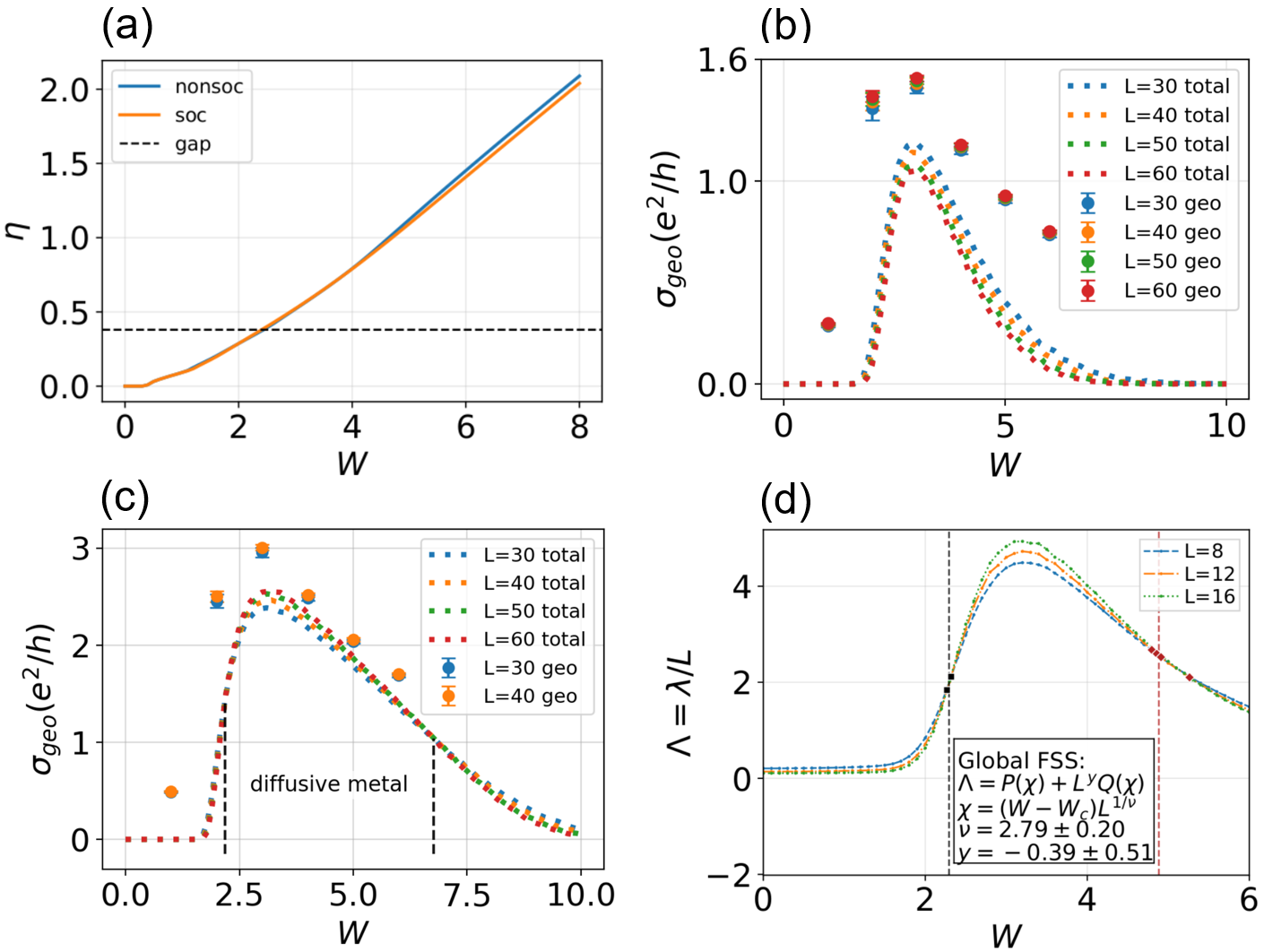}
\caption{Two terminal transport calculations and geometric conductance simulations. (a) Disorder induced spectral broadening \(\eta\) as a function of disorder strength \(W\) for both the SOC-free model and the Rashba-SOC model; the dashed horizontal line indicates the band gap \(\Delta\).
(b) SOC-free case: two-terminal conductance \(G\) and geometric conductance \(\sigma_{\rm geo}\) versus \(W\) for different system sizes \(L\); dashed curves denote the total conductance obtained from Green's function calculation, while symbols show \(\sigma_{\rm geo}\) obtained from eq.\ref{eq:trace_gL}.
(c) Rashba-SOC case: \(G\) and \(\sigma_{\rm geo}\) versus \(W\) for different \(L\), an intermediate diffusive-metal regime emerges.
(d) Rashba-SOC case: 1600 sample averaged renormalized localization length versus \(W\) for different \(L\); inset: finite size scaling near the first crossing yields the critical exponent \(\nu\).}\label{fig:ggeo}
\end{figure}

The scattering rate $\eta$ for the model without SOC and with Rashba SOC are shown in Fig.2 (a). In the weak disorder regime near the clean limit, $\eta$ remains zero. Upon increasing it turns on beyond a threshold and then grows monotonically with disorder strength. For \(W \sim 2\), the disorder-induced broadening becomes comparable to the gap, indicating that the system has entered the regime where quantum metric dominates\cite{PhysRevB.103.075415,PhysRevB.105.085154,PhysRevB.106.165133}.

In Fig. 2(b), both transport conductance and geometric conductivity of model without SOC are employed, and we show the conductance as a function of disorder strength. In the clean limit, the conductance is almost zero, corresponding to the insulating state caused by flat band localization. Notably, for \(W > 2\), the conductance increases rapidly to the order of \(e^2/h\). The peak occurs at \(W \sim 3.4\), after which the conductance starts to decrease, eventually entering the insulating phase of Anderson localization. The points shown in the figure represent the geometric conductivity, calculated using eq.\ref{eq:trace_gL}, which closely matches the transport conductance. As the system size increases, the conductance decreases slowly and monotonically. This behavior aligns with the predictions of single parameter scaling theory\cite{PhysRevLett.42.673}. Previous work on planar pyrochlore lattices with weak disorder has found critical behavior accompanied by multifractal features\cite{PhysRevB.82.104209,RevModPhys.80.1355}, where the system is neither Anderson localized nor spatially extended. Notably, a recent study demonstrated a strong enhancement of the space resolved quantum metric in a disordered setting and traced its origin to multifractal critical wave functions\cite{PhysRevLett.133.026002}, consistent with our findings. These demonstrate that the conductance observed under disorder is primarily due to the quantum metric, with no corresponding behavior in the clean limit.

In Fig. 2(c), we present the model that includes Rashba SOC. The resulting conductance versus disorder strength curves show a similar trend to that in Fig. 2(b), with flat band localization occurring at $W\le2.2$ and Anderson localization at $W\ge6.5$. However, the conductance values are larger, particularly in the intermediate region where two critical points emerge. In the range $2.2\le W\le6.5$, as the system size increases, the conductance grows, exhibiting metallic behavior. The geometric conductivity calculated using eq.\ref{eq:trace_gL} closely matches the transport conductance, and the observed larger conductance and metallic behavior can be attributed to the strengthening of the quantum metric due to Rashba SOC\cite{Sala2025QuantumMetricScience}. To investigate the metallic phase and extract the critical exponent, we calculate the sample averaged renormalized localization length\cite{PhysRevLett.47.1546,MacKinnon1983ZPhysB,Kramer1993RPP,PhysRevLett.82.382} and plot it as a function of disorder strength W for various system sizes in Fig. 2(d). The curves reveal two crossing points, signaling a two dimensional inverse Anderson transition\cite{PhysRevLett.96.126401}. We fit the critical exponent for the transition from flat band localization to the diffusive metal phase to be approximately 2.79, close to previous studies of 2D symplectic metal-insulator transitions\cite{HikamiLarkinNagaoka1980,PhysRevLett.89.256601,PhysRevB.70.035115}.

\begin{figure}[htb]
  \centering
  \includegraphics[width=\columnwidth]{fig3.png}
\caption{Schematic of the geometric percolation picture. Colored hemispheres represent effective potential valleys (``puddles'') whose local extent encodes the spread. Neighboring valleys are connected when quantum tunneling is sufficiently strong; in the illustration, the bright links highlight a system-spanning conducting path formed by connected valleys.}
\label{fig:perco}
\end{figure}

To further substantiate this picture, we introduce a geometric percolation model\cite{RevModPhys.45.574,RevModPhys.64.961,Stauffer1994PercolationTheory} rooted in real space quantum metric marker in Fig. 3. The gauge invariant Wannier spread $\Omega_I$, which quantifies the variance of the real space charge distribution\cite{PhysRevLett.82.370}, can be expressed in terms of a real space quantum metric marker g\cite{PhysRevB.56.12847,PhysRevB.110.075203},
$$
\Omega_I=\frac{V_{\text {cell }}}{\hbar^{D-2}} \operatorname{Tr} g_{\mu \nu}
$$
we thus associate each effective potential valley supporting a compact localized orbit with a site of an auxiliary lattice, and connect nearest-neighbor valleys by links that represent quantum tunneling between them\cite{PhysRevLett.83.3506,PhysRevB.61.16470}. When disorder is weak, $g_i$ is small and electronic trajectories remain trapped inside individual valleys, with only weak tunneling across the saddle region separating adjacent valleys. As disorder increases, the destructive interference confinement is progressively lifted, which enhances the gauge-invariant spread between neighboring valleys. Once this enhancement exceeds the local saddle-point threshold, two previously isolated trajectories coalesce: the electron can move freely from one valley to its neighbor, and the link connecting them becomes a perfect link\cite{PhysRevLett.94.156406}. In this coarse-grained network, joining two sites by a perfect link effectively merges them into a single site. Consequently, only when the perfect links percolate and form a system spanning cluster does the transmission become length independent, so that the critical percolation threshold is well defined and the conductance does not decay with system size. The local saddle-point threshold is model dependent, but the critical exponent is universal. This percolation of perfect links therefore provides a natural geometric mechanism for the localization–delocalization transition observed in transport. The resulting criticality is governed by quantum transport on this emergent network, where tunneling and interference determine the scaling behavior.

For model without SOC, we use the disorder strength \(W\), rather than the fraction of perfect links \(p\), as the control parameter for the percolation analysis. This choice is motivated by the fact that \(p(W)\) is typically analytic in the vicinity of \(W_c\) and is dominated by its linear term, widely used in the Interger Quantum Hall literature\cite{RevModPhys.67.357,Kramer2005PhysRep}. We define the site “spread” as the arithmetic square root of the local marker $g_i$ and choose $2\pi$ as the local saddle-point threshold. A nearest-neighbor bond ⟨ij⟩ is a perfect link if their combined spread exceeds the saddle threshold, and is otherwise closed. We generate \(N_s=200\) disorder realizations for each \((W,L)\), and define the spanning probability \(p_t\) \cite{PhysRev.133.A171,PhysRevLett.69.2670} as the fraction of samples that admit a conducting path connecting the left and right boundaries. The continuous evolution of \(p_t\) from \(0\) to \(1\) signals a percolation transition.
\begin{figure}[htb]
  \centering
  \includegraphics[width=\columnwidth]{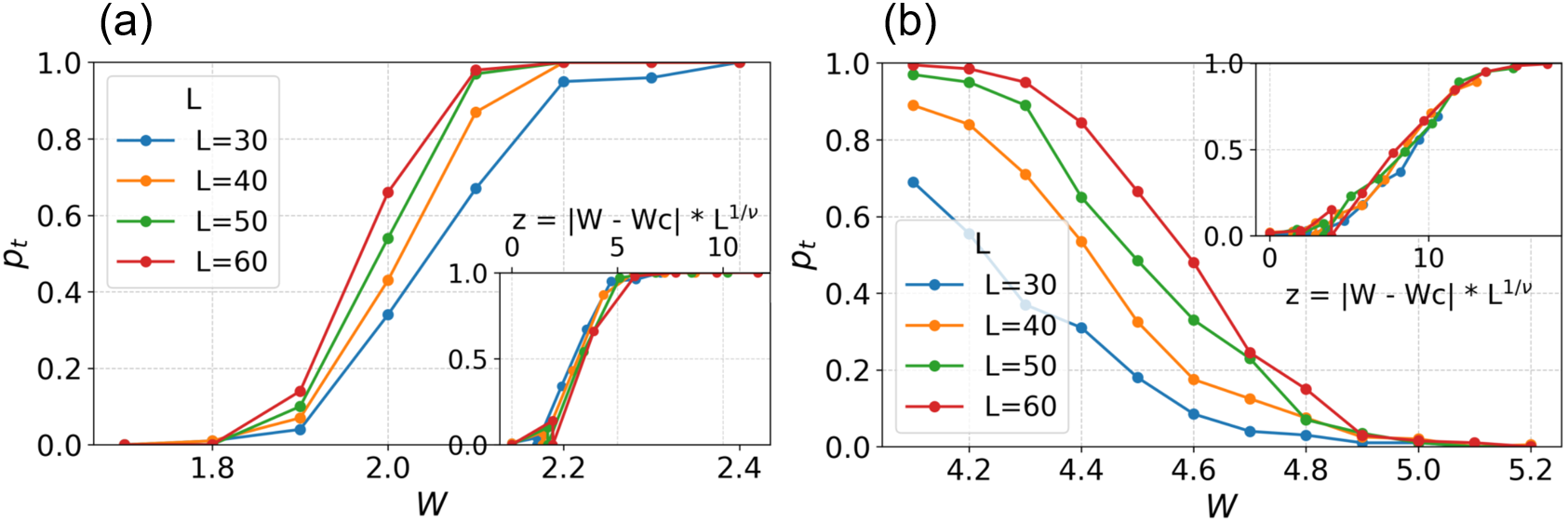}
\caption{Real space quantum metric controlled percolation transition. Spanning probability \(p_t\) versus disorder strength \(W\) for different system sizes \(L\), averaged over \(N_s=200\) disorder realizations. Insets show the same data plotted versus a rescaled horizontal axis ( \(z=(W-W_c)L^{1/\nu}\)) to demonstrate scaling collapse near criticality. (a) Window around the first threshold, \(W\in[1.7,\,2.4]\). (b) Window around the second threshold, \(W\in[4.1,\,5.2]\).}

\label{fig:pt}
\end{figure}

Fig. 4 shows \(p_t(W)\) for different system sizes \(L\). In Fig. 4(a) , around \(W\simeq 2\), \(p_t\) rises sharply from \(0\) toward \(1\) for all \(L\), indicating the onset of percolation. This threshold coincides with the transport defined boundary \(W_{c1}\) separating the flat-band localized regime from the quasi-metallic window. In the inset of Fig. 4(a), we perform a finite size scaling collapse using the dimensionless scaling variable $z = (W-W_{c})\,L^{1/\nu_1}$\cite{PhysRevB.71.125311}, and find that the data for different \(L\) collapse onto a single universal curve, providing strong evidence that the transition is governed by a diverging correlation length with exponent \(\nu_1\)=1.33, which is classical\cite{Shashkin1994PRB,Dunford2001PhysicaB}.

In Fig. 4(b), around \(W\simeq 4.5\), \(p_t\) decreases from \(1\) back to \(0\) as \(W\) increases, indicating that the system leaves the percolating critical regime and re-enters a localized phase. This second threshold matches the transport defined boundary \(W_{c2}\) between the quasi-metallic regime and the Anderson localized regime. The inset of Fig. 4(b) shows an analogous scaling collapse using $z = (W_{c}-W)\,L^{1/\nu_2}$ with exponent \(\nu_2\)=1.33 again yielding a single scaling curve. Together, these results support a percolation description of the entire evolution, encoded by the real space quantum metric marker.

In the presence of SOC, the transport can be captured by a quantum tunneling network model in the symplectic symmetry class\cite{PhysRevB.58.9627}.

In conclusion, we consider the two dimensional multi-flatband stub-pyrochlore model and identify a pronounced geometric linear conductivity adding disorder between the flat bands, which is directly tied to the real space quantum metric. This region is a critical state, upon incorporating SOC, the enhanced quantum metric transitions from the critical state to a diffusive metallic regime. We model the localization-delocalization transition with a classical percolation framework, where local "puddles" represent regions of spread, which govern electron transport. The critical state emerges as disorder weakens destructive interference, allowing larger puddles form a percolating cluster, leading to the delocalization transition. As disorder intensifies, puddle sizes shrink, eventually leading to Anderson localization. Therefore, the flat band delocalization can be understood as a classical percolation of quantum metric puddles.

 The authors would like to thank Tobias Holder, Zhida Song, Ziyue Qi for the helpful discussions.
This work was supported by the National Key R\&D Program of China (Grant No. 2023YFA1607400, 2022YFA1403800), the National Natural Science Foundation of China (Grant No.12274436, 11925408, 11921004), the Science Center of the National Natural Science Foundation of China (Grant No. 12188101). 

\bibliographystyle{unsrt}
\bibliography{ref.bib}

@article{PhysRevX.15.021091,
  title = {Flat-Band (De)localization Emulated with a Superconducting Qubit Array},
  author = {Rosen, Ilan T. and Muschinske, Sarah and Barrett, Cora N. and Rower, David A. and Das, Rabindra and Kim, David K. and Niedzielski, Bethany M. and Schuldt, Meghan and Serniak, Kyle and Schwartz, Mollie E. and Yoder, Jonilyn L. and Grover, Jeffrey A. and Oliver, William D.},
  journal = {Phys. Rev. X},
  volume = {15},
  issue = {2},
  pages = {021091},
  numpages = {17},
  year = {2025},
  month = {Jun},
  publisher = {American Physical Society},
  doi = {10.1103/PhysRevX.15.021091},
  url = {https://link.aps.org/doi/10.1103/PhysRevX.15.021091}
}

@article{PhysRevB.103.075415,
  title = {Quantum transport in flat bands and supermetallicity},
  author = {Bouzerar, G. and Mayou, D.},
  journal = {Phys. Rev. B},
  volume = {103},
  issue = {7},
  pages = {075415},
  numpages = {5},
  year = {2021},
  month = {Feb},
  publisher = {American Physical Society},
  doi = {10.1103/PhysRevB.103.075415},
  url = {https://link.aps.org/doi/10.1103/PhysRevB.103.075415}
}

@article{PhysRevB.105.085154,
  title = {Bound on resistivity in flat-band materials due to the quantum metric},
  author = {Mitscherling, Johannes and Holder, Tobias},
  journal = {Phys. Rev. B},
  volume = {105},
  issue = {8},
  pages = {085154},
  numpages = {11},
  year = {2022},
  month = {Feb},
  publisher = {American Physical Society},
  doi = {10.1103/PhysRevB.105.085154},
  url = {https://link.aps.org/doi/10.1103/PhysRevB.105.085154}
}

@article{PhysRevB.109.035303,
  title = {Disorder-induced phase transitions in three-dimensional chiral second-order topological insulator},
  author = {Shen, Yedi and Li, Zeyu and Niu, Qian and Qiao, Zhenhua},
  journal = {Phys. Rev. B},
  volume = {109},
  issue = {3},
  pages = {035303},
  numpages = {6},
  year = {2024},
  month = {Jan},
  publisher = {American Physical Society},
  doi = {10.1103/PhysRevB.109.035303},
  url = {https://link.aps.org/doi/10.1103/PhysRevB.109.035303}
}

@article{PhysRevLett.103.196805,
  title = {Theory of the Topological Anderson Insulator},
  author = {Groth, C. W. and Wimmer, M. and Akhmerov, A. R. and Tworzyd\l{}o, J. and Beenakker, C. W. J.},
  journal = {Phys. Rev. Lett.},
  volume = {103},
  issue = {19},
  pages = {196805},
  numpages = {4},
  year = {2009},
  month = {Nov},
  publisher = {American Physical Society},
  doi = {10.1103/PhysRevLett.103.196805},
  url = {https://link.aps.org/doi/10.1103/PhysRevLett.103.196805}
}

@article{Provost1980Riemannian,
  author  = {Provost, J. P. and Vall{\'e}e, G.},
  title   = {Riemannian structure on manifolds of quantum states},
  journal = {Communications in Mathematical Physics},
  volume  = {76},
  number  = {3},
  pages   = {289--301},
  year    = {1980},
  month   = {Sep},
  doi     = {10.1007/BF02193559},
  url     = {https://doi.org/10.1007/BF02193559}
}

@article{Berry1984Quantal,
  author  = {Berry, M. V.},
  title   = {Quantal phase factors accompanying adiabatic changes},
  journal = {Proceedings of the Royal Society of London. Series A, Mathematical and Physical Sciences},
  volume  = {392},
  number  = {1802},
  pages   = {45--57},
  year    = {1984},
  month   = {Mar},
  doi     = {10.1098/rspa.1984.0023},
  url     = {https://doi.org/10.1098/rspa.1984.0023}
}

@article{Simon1983Holonomy,
  author  = {Simon, Barry},
  title   = {Holonomy, the Quantum Adiabatic Theorem, and Berry's Phase},
  journal = {Physical Review Letters},
  volume  = {51},
  number  = {24},
  pages   = {2167--2170},
  year    = {1983},
  doi     = {10.1103/PhysRevLett.51.2167},
  url     = {https://doi.org/10.1103/PhysRevLett.51.2167}
}

@article{Aharonov1987Phase,
  author  = {Aharonov, Yakir and Anandan, Jeeva},
  title   = {Phase change during a cyclic quantum evolution},
  journal = {Physical Review Letters},
  volume  = {58},
  number  = {16},
  pages   = {1593--1596},
  year    = {1987},
  month   = {Apr},
  doi     = {10.1103/PhysRevLett.58.1593},
  url     = {https://doi.org/10.1103/PhysRevLett.58.1593}
}

@article{PhysRevB.106.165133,
  title = {Nontrivial quantum geometry of degenerate flat bands},
  author = {Mera, Bruno and Mitscherling, Johannes},
  journal = {Phys. Rev. B},
  volume = {106},
  issue = {16},
  pages = {165133},
  numpages = {8},
  year = {2022},
  month = {Oct},
  publisher = {American Physical Society},
  doi = {10.1103/PhysRevB.106.165133},
  url = {https://link.aps.org/doi/10.1103/PhysRevB.106.165133}
}

@article{PhysRevB.82.104209,
  title = {Anderson localization in tight-binding models with flat bands},
  author = {Chalker, J. T. and Pickles, T. S. and Shukla, Pragya},
  journal = {Phys. Rev. B},
  volume = {82},
  issue = {10},
  pages = {104209},
  numpages = {5},
  year = {2010},
  month = {Sep},
  publisher = {American Physical Society},
  doi = {10.1103/PhysRevB.82.104209},
  url = {https://link.aps.org/doi/10.1103/PhysRevB.82.104209}
}

@article{PhysRevLett.133.026002,
  title = {Enhanced Quantum Metric due to Vacancies in Graphene},
  author = {Marsal, Quentin and Black-Schaffer, Annica M.},
  journal = {Phys. Rev. Lett.},
  volume = {133},
  issue = {2},
  pages = {026002},
  numpages = {6},
  year = {2024},
  month = {Jul},
  publisher = {American Physical Society},
  doi = {10.1103/PhysRevLett.133.026002},
  url = {https://link.aps.org/doi/10.1103/PhysRevLett.133.026002}
}

@article{dzqm-9kwj,
  title = {Family of Aperiodic Tilings with Tunable Quantum Geometric Tensor},
  author = {Roche Carrasco, Hector and Schirmann, Justin and Mordret, Aurelien and Grushin, Adolfo G.},
  journal = {Phys. Rev. Lett.},
  volume = {135},
  issue = {23},
  pages = {236603},
  numpages = {11},
  year = {2025},
  month = {Dec},
  publisher = {American Physical Society},
  doi = {10.1103/dzqm-9kwj},
  url = {https://link.aps.org/doi/10.1103/dzqm-9kwj}
}

@article{Shashkin1994PRB,
  author  = {Shashkin, A. A. and Dolgopolov, V. T. and Kravchenko, G. V.},
  title   = {Insulating phases in a two-dimensional electron system of high-mobility Si MOSFET's},
  journal = {Phys. Rev. B},
  volume  = {49},
  pages   = {14486},
  year    = {1994},
  doi     = {10.1103/PhysRevB.49.14486}
}

@article{Dunford2001PhysicaB,
  author  = {Dunford, R. B. and Griffin, N. and Phillips, P. J. and Whall, T. E.},
  title   = {Weak localisation and inter-quantum Hall effect transitions in a 2D Si/SiGe hole system},
  journal = {Physica B: Condensed Matter},
  volume  = {298},
  number  = {1--4},
  pages   = {496--500},
  year    = {2001},
  doi     = {10.1016/S0921-4526(01)00370-2}
}

@article{RevModPhys.67.357,
  title = {Scaling theory of the integer quantum Hall effect},
  author = {Huckestein, Bodo},
  journal = {Rev. Mod. Phys.},
  volume = {67},
  issue = {2},
  pages = {357--396},
  numpages = {0},
  year = {1995},
  month = {Apr},
  publisher = {American Physical Society},
  doi = {10.1103/RevModPhys.67.357},
  url = {https://link.aps.org/doi/10.1103/RevModPhys.67.357}
}

@article{PhysRevLett.69.2670,
  title = {Spanning probability in 2D percolation},
  author = {Ziff, Robert M.},
  journal = {Phys. Rev. Lett.},
  volume = {69},
  issue = {18},
  pages = {2670--2673},
  numpages = {0},
  year = {1992},
  month = {Nov},
  publisher = {American Physical Society},
  doi = {10.1103/PhysRevLett.69.2670},
  url = {https://link.aps.org/doi/10.1103/PhysRevLett.69.2670}
}

@article{PhysRevLett.89.256601,
  title = {Anderson Transition in Two-Dimensional Systems with Spin-Orbit Coupling},
  author = {Asada, Yoichi and Slevin, Keith and Ohtsuki, Tomi},
  journal = {Phys. Rev. Lett.},
  volume = {89},
  issue = {25},
  pages = {256601},
  numpages = {4},
  year = {2002},
  month = {Dec},
  publisher = {American Physical Society},
  doi = {10.1103/PhysRevLett.89.256601},
  url = {https://link.aps.org/doi/10.1103/PhysRevLett.89.256601}
}

@article{PhysRevB.58.9627,
  title = {Two-dimensional random-network model with symplectic symmetry},
  author = {Minakuchi, Kei},
  journal = {Phys. Rev. B},
  volume = {58},
  issue = {15},
  pages = {9627--9630},
  numpages = {0},
  year = {1998},
  month = {Oct},
  publisher = {American Physical Society},
  doi = {10.1103/PhysRevB.58.9627},
  url = {https://link.aps.org/doi/10.1103/PhysRevB.58.9627}
}

@article{PhysRevLett.96.126401,
  title = {Inverse Anderson Transition Caused by Flatbands},
  author = {Goda, Masaki and Nishino, Shinya and Matsuda, Hiroki},
  journal = {Phys. Rev. Lett.},
  volume = {96},
  issue = {12},
  pages = {126401},
  numpages = {4},
  year = {2006},
  month = {Mar},
  publisher = {American Physical Society},
  doi = {10.1103/PhysRevLett.96.126401},
  url = {https://link.aps.org/doi/10.1103/PhysRevLett.96.126401}
}

@article{Kramer2005PhysRep,
  author  = {Kramer, B. and Ohtsuki, T. and Kettemann, S.},
  title   = {Random Network Models and Quantum Phase Transitions in Two Dimensions},
  journal = {Physics Reports},
  volume  = {417},
  number  = {5--6},
  pages   = {211--342},
  year    = {2005},
  doi     = {10.1016/j.physrep.2005.07.001}
}

@misc{Chau2024DisorderInducedDiffusion,
  author       = {Chau, Chun Wang and Xiang, Tian and Chen, Shuai A. and Law, K. T.},
  title        = {Disorder-induced diffusion transport in flat-band systems with quantum metric},
  howpublished = {arXiv:2412.19056},
  year         = {2024},
  eprint       = {2412.19056},
  archivePrefix= {arXiv},
  primaryClass = {cond-mat.mes-hall},
  doi          = {10.48550/arXiv.2412.19056}
}

@article{PhysRevB.110.075203,
  title = {Detecting the spread of valence-band Wannier functions by optical sum rules},
  author = {C\'ardenas-Castillo, Luis F. and Zhang, Shuai and Freire, Fernando L. and Kochan, Denis and Chen, Wei},
  journal = {Phys. Rev. B},
  volume = {110},
  issue = {7},
  pages = {075203},
  numpages = {12},
  year = {2024},
  month = {Aug},
  publisher = {American Physical Society},
  doi = {10.1103/PhysRevB.110.075203},
  url = {https://link.aps.org/doi/10.1103/PhysRevB.110.075203}
}

@article{10.1093/nsr/nwae334,
    author = {Liu, Tianyu and Qiang, Xiao-Bin and Lu, Hai-Zhou and Xie, X C},
    title = {Quantum geometry in condensed matter},
    journal = {National Science Review},
    volume = {12},
    number = {3},
    pages = {nwae334},
    year = {2024},
    month = {09},
    issn = {2095-5138},
    doi = {10.1093/nsr/nwae334},
    url = {https://doi.org/10.1093/nsr/nwae334},
    eprint = {https://academic.oup.com/nsr/article-pdf/12/3/nwae334/59204701/nwae334.pdf},
}

@article{PhysRevB.71.125311,
  title = {Quantum Hall criticality, superconductor-insulator transition, and quantum percolation},
  author = {Dubi, Yonatan and Meir, Yigal and Avishai, Yshai},
  journal = {Phys. Rev. B},
  volume = {71},
  issue = {12},
  pages = {125311},
  numpages = {4},
  year = {2005},
  month = {Mar},
  publisher = {American Physical Society},
  doi = {10.1103/PhysRevB.71.125311},
  url = {https://link.aps.org/doi/10.1103/PhysRevB.71.125311}
}

@article{PhysRevLett.94.156406,
  title = {Unifying Model for Several Classes of Two-Dimensional Phase Transition},
  author = {Dubi, Yonatan and Meir, Yigal and Avishai, Yshai},
  journal = {Phys. Rev. Lett.},
  volume = {94},
  issue = {15},
  pages = {156406},
  numpages = {4},
  year = {2005},
  month = {Apr},
  publisher = {American Physical Society},
  doi = {10.1103/PhysRevLett.94.156406},
  url = {https://link.aps.org/doi/10.1103/PhysRevLett.94.156406}
}

@article{PhysRevLett.132.026002,
  title = {Ginzburg-Landau Theory of Flat-Band Superconductors with Quantum Metric},
  author = {Chen, Shuai A. and Law, K. T.},
  journal = {Phys. Rev. Lett.},
  volume = {132},
  issue = {2},
  pages = {026002},
  numpages = {7},
  year = {2024},
  month = {Jan},
  publisher = {American Physical Society},
  doi = {10.1103/PhysRevLett.132.026002},
  url = {https://link.aps.org/doi/10.1103/PhysRevLett.132.026002}
}

@article{Yu2025QuantumGeometryNPJQM,
  author  = {Yu, Jiabin and Bernevig, B. Andrei and Queiroz, Raquel and Rossi, Enrico and T{\"o}rm{\"a}, P{\"a}ivi and Yang, Bohm-Jung},
  title   = {Quantum geometry in quantum materials},
  journal = {npj Quantum Materials},
  volume  = {10},
  pages   = {101},
  year    = {2025},
  doi     = {10.1038/s41535-025-00801-3}
}

@misc{Hou2025FateDisorderTBG,
  author       = {Hou, Zhe and Li, Hailong and Yan, Qing and Li, Yu-Hang and Jiang, Hua},
  title        = {The fate of disorder in twisted bilayer graphene near the magic angle},
  howpublished = {arXiv:2510.14567v1},
  year         = {2025},
  eprint       = {2510.14567},
  archivePrefix= {arXiv},
  primaryClass = {cond-mat.mes-hall},
  doi          = {10.48550/arXiv.2510.14567}
}

@article{Torma2022NatRevPhysQuantumGeometry,
  author  = {T{\"o}rm{\"a}, P{\"a}ivi and Peotta, Sebastiano and Bernevig, Bogdan A.},
  title   = {Superconductivity, superfluidity and quantum geometry in twisted multilayer systems},
  journal = {Nature Reviews Physics},
  volume  = {4},
  number  = {8},
  pages   = {528--542},
  year    = {2022},
  doi     = {10.1038/s42254-022-00466-y}
}

@article{RevModPhys.82.1959,
  title = {Berry phase effects on electronic properties},
  author = {Xiao, Di and Chang, Ming-Che and Niu, Qian},
  journal = {Rev. Mod. Phys.},
  volume = {82},
  issue = {3},
  pages = {1959--2007},
  numpages = {0},
  year = {2010},
  month = {Jul},
  publisher = {American Physical Society},
  doi = {10.1103/RevModPhys.82.1959},
  url = {https://link.aps.org/doi/10.1103/RevModPhys.82.1959}
}

@article{PhysRevB.90.165139,
  title = {Band geometry of fractional topological insulators},
  author = {Roy, Rahul},
  journal = {Phys. Rev. B},
  volume = {90},
  issue = {16},
  pages = {165139},
  numpages = {7},
  year = {2014},
  month = {Oct},
  publisher = {American Physical Society},
  doi = {10.1103/PhysRevB.90.165139},
  url = {https://link.aps.org/doi/10.1103/PhysRevB.90.165139}
}

@article{PhysRevB.104.045104,
  title = {K\"ahler geometry and Chern insulators: Relations between topology and the quantum metric},
  author = {Mera, Bruno and Ozawa, Tomoki},
  journal = {Phys. Rev. B},
  volume = {104},
  issue = {4},
  pages = {045104},
  numpages = {13},
  year = {2021},
  month = {Jul},
  publisher = {American Physical Society},
  doi = {10.1103/PhysRevB.104.045104},
  url = {https://link.aps.org/doi/10.1103/PhysRevB.104.045104}
}

@article{PhysRevB.108.205144,
  title = {Vortexability: A unifying criterion for ideal fractional Chern insulators},
  author = {Ledwith, Patrick J. and Vishwanath, Ashvin and Parker, Daniel E.},
  journal = {Phys. Rev. B},
  volume = {108},
  issue = {20},
  pages = {205144},
  numpages = {20},
  year = {2023},
  month = {Nov},
  publisher = {American Physical Society},
  doi = {10.1103/PhysRevB.108.205144},
  url = {https://link.aps.org/doi/10.1103/PhysRevB.108.205144}
}

@article{PhysRevB.102.201112,
  title = {Superconductivity, pseudogap, and phase separation in topological flat bands},
  author = {Hofmann, Johannes S. and Berg, Erez and Chowdhury, Debanjan},
  journal = {Phys. Rev. B},
  volume = {102},
  issue = {20},
  pages = {201112},
  numpages = {6},
  year = {2020},
  month = {Nov},
  publisher = {American Physical Society},
  doi = {10.1103/PhysRevB.102.201112},
  url = {https://link.aps.org/doi/10.1103/PhysRevB.102.201112}
}

@article{Peotta2015NatCommun,
  author  = {Peotta, Sebastiano and T{\"o}rm{\"a}, P{\"a}ivi},
  title   = {Superfluidity in topologically nontrivial flat bands},
  journal = {Nature Communications},
  volume  = {6},
  pages   = {8944},
  year    = {2015},
  doi     = {10.1038/ncomms9944}
}

@article{PhysRevLett.129.220403,
  title = {Aharonov-Bohm Caging and Inverse Anderson Transition in Ultracold Atoms},
  author = {Li, Hang and Dong, Zhaoli and Longhi, Stefano and Liang, Qian and Xie, Dizhou and Yan, Bo},
  journal = {Phys. Rev. Lett.},
  volume = {129},
  issue = {22},
  pages = {220403},
  numpages = {6},
  year = {2022},
  month = {Nov},
  publisher = {American Physical Society},
  doi = {10.1103/PhysRevLett.129.220403},
  url = {https://link.aps.org/doi/10.1103/PhysRevLett.129.220403}
}

@article{PhysRevLett.130.206401,
  title = {Non-Abelian Inverse Anderson Transitions},
  author = {Zhang, Weixuan and Wang, Haiteng and Sun, Houjun and Zhang, Xiangdong},
  journal = {Phys. Rev. Lett.},
  volume = {130},
  issue = {20},
  pages = {206401},
  numpages = {6},
  year = {2023},
  month = {May},
  publisher = {American Physical Society},
  doi = {10.1103/PhysRevLett.130.206401},
  url = {https://link.aps.org/doi/10.1103/PhysRevLett.130.206401}
}

@article{PhysRevB.88.224203,
  title = {Flat band states: Disorder and nonlinearity},
  author = {Leykam, Daniel and Flach, Sergej and Bahat-Treidel, Omri and Desyatnikov, Anton S.},
  journal = {Phys. Rev. B},
  volume = {88},
  issue = {22},
  pages = {224203},
  numpages = {6},
  year = {2013},
  month = {Dec},
  publisher = {American Physical Society},
  doi = {10.1103/PhysRevB.88.224203},
  url = {https://link.aps.org/doi/10.1103/PhysRevB.88.224203}
}

@article{PhysRevLett.128.087002,
  title = {Superfluid Weight Bounds from Symmetry and Quantum Geometry in Flat Bands},
  author = {Herzog-Arbeitman, Jonah and Peri, Valerio and Schindler, Frank and Huber, Sebastian D. and Bernevig, B. Andrei},
  journal = {Phys. Rev. Lett.},
  volume = {128},
  issue = {8},
  pages = {087002},
  numpages = {8},
  year = {2022},
  month = {Feb},
  publisher = {American Physical Society},
  doi = {10.1103/PhysRevLett.128.087002},
  url = {https://link.aps.org/doi/10.1103/PhysRevLett.128.087002}
}

@article{Resta2011EurPhysJ,
  author  = {Resta, R.},
  title   = {The insulating state of matter: A geometrical theory},
  journal = {European Physical Journal B},
  volume  = {79},
  pages   = {121},
  year    = {2011},
  doi     = {10.1140/epjb/e2011-10799-5}
}

@article{PhysRevB.111.134201,
  title = {Scaling of the integrated quantum metric in disordered topological phases},
  author = {Romeral, Jorge Mart\'{\i}nez and Cummings, Aron W. and Roche, Stephan},
  journal = {Phys. Rev. B},
  volume = {111},
  issue = {13},
  pages = {134201},
  numpages = {8},
  year = {2025},
  month = {Apr},
  publisher = {American Physical Society},
  doi = {10.1103/PhysRevB.111.134201},
  url = {https://link.aps.org/doi/10.1103/PhysRevB.111.134201}
}

@article{PhysRevLett.132.026301,
  title = {Unification of Nonlinear Anomalous Hall Effect and Nonreciprocal Magnetoresistance in Metals by the Quantum Geometry},
  author = {Kaplan, Daniel and Holder, Tobias and Yan, Binghai},
  journal = {Phys. Rev. Lett.},
  volume = {132},
  issue = {2},
  pages = {026301},
  numpages = {7},
  year = {2024},
  month = {Jan},
  publisher = {American Physical Society},
  doi = {10.1103/PhysRevLett.132.026301},
  url = {https://link.aps.org/doi/10.1103/PhysRevLett.132.026301}
}

@article{PhysRevLett.127.277201,
  title = {Intrinsic Nonlinear Hall Effect in Antiferromagnetic Tetragonal CuMnAs},
  author = {Wang, Chong and Gao, Yang and Xiao, Di},
  journal = {Phys. Rev. Lett.},
  volume = {127},
  issue = {27},
  pages = {277201},
  numpages = {6},
  year = {2021},
  month = {Dec},
  publisher = {American Physical Society},
  doi = {10.1103/PhysRevLett.127.277201},
  url = {https://link.aps.org/doi/10.1103/PhysRevLett.127.277201}
}

@article{PhysRevLett.127.277202,
  title = {Intrinsic Second-Order Anomalous Hall Effect and Its Application in Compensated Antiferromagnets},
  author = {Liu, Huiying and Zhao, Jianzhou and Huang, Yue-Xin and Wu, Weikang and Sheng, Xian-Lei and Xiao, Cong and Yang, Shengyuan A.},
  journal = {Phys. Rev. Lett.},
  volume = {127},
  issue = {27},
  pages = {277202},
  numpages = {6},
  year = {2021},
  month = {Dec},
  publisher = {American Physical Society},
  doi = {10.1103/PhysRevLett.127.277202},
  url = {https://link.aps.org/doi/10.1103/PhysRevLett.127.277202}
}

@article{Leykam2017EPJB,
  author  = {Leykam, Daniel and Bodyfelt, Joshua D. and Desyatnikov, Anton S. and Flach, Sergej},
  title   = {Localization of weakly disordered flat band states},
  journal = {European Physical Journal B},
  volume  = {90},
  number  = {1},
  pages   = {1--12},
  year    = {2017},
  doi     = {10.1140/epjb/e2016-70551-2}
}

@article{Gao2023QuantumMetricScience,
  author  = {Gao, Anyuan and Liu, Yu-Fei and Qiu, Jian-Xiang and Ghosh, Barun and Trevisan, Tha\'{i}s V. and Onishi, Yugo and Hu, Chaowei and Qian, Tiema and Tien, Hung-Ju and Chen, Shao-Wen and Huang, Mengqi and B\'{e}rub\'{e}, Damien and Li, Houchen and Tzschaschel, Christian and Dinh, Thao and Sun, Zhe and Ho, Sheng-Chin and Lien, Shang-Wei and Singh, Bahadur and Watanabe, Kenji and Taniguchi, Takashi and Bell, David C. and Lin, Hsin and Chang, Tay-Rong and Du, Chunhui Rita and Bansil, Arun and Fu, Liang and Ni, Ni and Orth, Peter P. and Ma, Qiong and Xu, Su-Yang},
  title   = {Quantum metric nonlinear Hall effect in a topological antiferromagnetic heterostructure},
  journal = {Science},
  volume  = {381},
  number  = {6654},
  pages   = {181--186},
  year    = {2023},
  doi     = {10.1126/science.adf1506}
}

@article{Wang2023QuantumMetricNature,
  author  = {Wang, Naizhou and Kaplan, Daniel and Zhang, Zhaowei and Holder, Tobias and Cao, Ning and Wang, Aifeng and Zhou, Xiaoyuan and Zhou, Feifei and Jiang, Zhengzhi and Zhang, Chusheng and Ru, Shihao and Cai, Hongbing and Watanabe, Kenji and Taniguchi, Takashi and Yan, Binghai and Gao, Weibo},
  title   = {Quantum-metric-induced nonlinear transport in a topological antiferromagnet},
  journal = {Nature},
  volume  = {621},
  number  = {7979},
  pages   = {487--492},
  year    = {2023},
  doi     = {10.1038/s41586-023-06363-3}
}

@article{RevModPhys.64.961,
  title = {Percolation, statistical topography, and transport in random media},
  author = {Isichenko, M. B.},
  journal = {Rev. Mod. Phys.},
  volume = {64},
  issue = {4},
  pages = {961--1043},
  numpages = {0},
  year = {1992},
  month = {Oct},
  publisher = {American Physical Society},
  doi = {10.1103/RevModPhys.64.961},
  url = {https://link.aps.org/doi/10.1103/RevModPhys.64.961}
}

@article{RevModPhys.80.1355,
  title = {Anderson transitions},
  author = {Evers, Ferdinand and Mirlin, Alexander D.},
  journal = {Rev. Mod. Phys.},
  volume = {80},
  issue = {4},
  pages = {1355--1417},
  numpages = {0},
  year = {2008},
  month = {Oct},
  publisher = {American Physical Society},
  doi = {10.1103/RevModPhys.80.1355},
  url = {https://link.aps.org/doi/10.1103/RevModPhys.80.1355}
}

@article{PhysRevLett.42.673,
  title = {Scaling Theory of Localization: Absence of Quantum Diffusion in Two Dimensions},
  author = {Abrahams, E. and Anderson, P. W. and Licciardello, D. C. and Ramakrishnan, T. V.},
  journal = {Phys. Rev. Lett.},
  volume = {42},
  issue = {10},
  pages = {673--676},
  numpages = {0},
  year = {1979},
  month = {Mar},
  publisher = {American Physical Society},
  doi = {10.1103/PhysRevLett.42.673},
  url = {https://link.aps.org/doi/10.1103/PhysRevLett.42.673}
}

@article{PhysRevB.56.12847,
  title = {Maximally localized generalized Wannier functions for composite energy bands},
  author = {Marzari, Nicola and Vanderbilt, David},
  journal = {Phys. Rev. B},
  volume = {56},
  issue = {20},
  pages = {12847--12865},
  numpages = {0},
  year = {1997},
  month = {Nov},
  publisher = {American Physical Society},
  doi = {10.1103/PhysRevB.56.12847},
  url = {https://link.aps.org/doi/10.1103/PhysRevB.56.12847}
}

@article{PhysRev.133.A171,
  title = {Theory of the Insulating State},
  author = {Kohn, Walter},
  journal = {Phys. Rev.},
  volume = {133},
  issue = {1A},
  pages = {A171--A181},
  numpages = {0},
  year = {1964},
  month = {Jan},
  publisher = {American Physical Society},
  doi = {10.1103/PhysRev.133.A171},
  url = {https://link.aps.org/doi/10.1103/PhysRev.133.A171}
}

@article{PhysRevLett.82.370,
  title = {Electron Localization in the Insulating State},
  author = {Resta, Raffaele and Sorella, Sandro},
  journal = {Phys. Rev. Lett.},
  volume = {82},
  issue = {2},
  pages = {370--373},
  numpages = {0},
  year = {1999},
  month = {Jan},
  publisher = {American Physical Society},
  doi = {10.1103/PhysRevLett.82.370},
  url = {https://link.aps.org/doi/10.1103/PhysRevLett.82.370}
}

@article{PhysRevB.84.241106,
  title = {Mapping topological order in coordinate space},
  author = {Bianco, Raffaello and Resta, Raffaele},
  journal = {Phys. Rev. B},
  volume = {84},
  issue = {24},
  pages = {241106},
  numpages = {4},
  year = {2011},
  month = {Dec},
  publisher = {American Physical Society},
  doi = {10.1103/PhysRevB.84.241106},
  url = {https://link.aps.org/doi/10.1103/PhysRevB.84.241106}
}

@book{Stauffer1994PercolationTheory,
  author    = {Dietrich Stauffer and Amnon Aharony},
  title     = {Introduction to Percolation Theory},
  edition   = {2nd},
  publisher = {Taylor \& Francis},
  year      = {1994}
}

@article{Rashba1960ExtremumLoop,
  author    = {E. I. Rashba},
  title     = {Properties of semiconductors with an extremum loop. I.},
  journal   = {Sov. Phys. -- Solid State},
  volume    = {2},
  pages     = {1109},
  year      = {1960}
}

@article{BychkovRashba1984,
  author    = {Y. A. Bychkov and E. I. Rashba},
  title     = {Properties of a 2D electron gas with lifted spectral degeneracy},
  journal   = {JETP Lett.},
  volume    = {39},
  pages     = {78--81},
  year      = {1984}
}

@article{HikamiLarkinNagaoka1980,
  author    = {S. Hikami and A. I. Larkin and Y. Nagaoka},
  title     = {Spin--Orbit Interaction and Magnetoresistance in the Two Dimensional Random System},
  journal   = {Prog. Theor. Phys.},
  volume    = {63},
  pages     = {707--710},
  year      = {1980}
}

@article{PhysRev.109.1492,
  title = {Absence of Diffusion in Certain Random Lattices},
  author = {Anderson, P. W.},
  journal = {Phys. Rev.},
  volume = {109},
  issue = {5},
  pages = {1492--1505},
  numpages = {0},
  year = {1958},
  month = {Mar},
  publisher = {American Physical Society},
  doi = {10.1103/PhysRev.109.1492},
  url = {https://link.aps.org/doi/10.1103/PhysRev.109.1492}
}

@article{PhysRevLett.115.216806,
  title = {Quantum Nonlinear Hall Effect Induced by Berry Curvature Dipole in Time-Reversal Invariant Materials},
  author = {Sodemann, Inti and Fu, Liang},
  journal = {Phys. Rev. Lett.},
  volume = {115},
  issue = {21},
  pages = {216806},
  numpages = {5},
  year = {2015},
  month = {Nov},
  publisher = {American Physical Society},
  doi = {10.1103/PhysRevLett.115.216806},
  url = {https://link.aps.org/doi/10.1103/PhysRevLett.115.216806}
}

@article{PhysRevLett.113.236403,
  title = {Flatbands under Correlated Perturbations},
  author = {Bodyfelt, Joshua D. and Leykam, Daniel and Danieli, Carlo and Yu, Xiaoquan and Flach, Sergej},
  journal = {Phys. Rev. Lett.},
  volume = {113},
  issue = {23},
  pages = {236403},
  numpages = {5},
  year = {2014},
  month = {Dec},
  publisher = {American Physical Society},
  doi = {10.1103/PhysRevLett.113.236403},
  url = {https://link.aps.org/doi/10.1103/PhysRevLett.113.236403}
}
\newpage
\clearpage

       \beginsupplement{}
        \setcounter{section}{0}
        \renewcommand{\thesubsection}{S\arabic{subsection}}
        \renewcommand{\thesubsubsection}{\Alph{subsubsection}}
        \titleformat{\section}[hang]{\bf\centering\Large}{\section}{1.0em}{}[]
        \titleformat{\subsection}[hang]{\bf\large}{\thesubsection}{0.5em}{}[]
        \titleformat{\subsubsection}[hang]{\bf\normalsize}{\thesubsubsection}{0.5em}{}[]

\clearpage
 
\end{document}